\documentclass[12pt]{ecoc02}

\usepackage{amsmath,amssymb,epsf}

\usepackage{mathptmx,graphicx}
\usepackage{algorithm}
\usepackage{algpseudocode}
\usepackage{pdfpages,cite,hyperref}

\begin{document}

\title{PPF -- A Parallel Particle Filtering Library}

\markboth{\"Omer Demirel, Ihor Smal, Wiro J.~Niessen, Erik Meijering and Ivo F.~Sbalzarini}{PPF- A Parallel Particle Filtering Library}

\author{\"Omer Demirel$^{*}$, Ihor Smal$^{\dagger}$, Wiro J.~Niessen$^{\dagger}$, Erik Meijering$^{\dagger}$ and Ivo F.~Sbalzarini$^{*}$}

\address{$^{*}$\textit{MOSAIC Group, Center of Systems Biology Dresden (CSBD),
Max Planck Institute of Molecular Cell Biology and Genetics,
Pfotenhauerstr. 108, 01307 Dresden, Germany. email: \{demirel,ivos\}@mpi-cbg.de}\\
$^{\dagger}$\textit{Biomedical Imaging Group Rotterdam, Departments of Medical Informatics and Radiology,  
Erasmus MC -- University Medical Center Rotterdam, Rotterdam, The Netherlands. email: \{i.smal,w.niessen\}@erasmusmc.nl, meijering@imagescience.org}}


\begin{abstract}
We present the parallel particle filtering (\texttt{PPF}) software library, which enables hybrid shared-memory/distributed-memory parallelization of particle filtering (PF) algorithms combining the Message Passing Interface (MPI) with multithreading for multi-level parallelism. The library is implemented in Java and relies on OpenMPI's Java bindings for inter-process communication. It includes dynamic load balancing, multi-thread balancing, and several algorithmic improvements for PF, such as input-space domain decomposition. The \texttt{PPF} library hides the difficulties of efficient parallel programming of PF algorithms and provides application developers with a tool for parallel implementation of PF methods. We demonstrate the capabilities of the \texttt{PPF} library using two distributed PF algorithms in two scenarios with different numbers of particles. The \texttt{PPF} library runs a 38 million particle problem, corresponding to more than 1.86\,GB of particle data, on 192 cores with 67\% parallel efficiency. 
\end{abstract}

\twocolumn

\maketitle

\section{Introduction}

Particle filters (PFs) are popular algorithms for tracking (multiple) targets under non-linear, non-Gaussian dynamics and observation models. From econometrics~\cite{flury2011bayesian} to robotics~\cite{tamimi2006localization} and sports~\cite{breitenstein2009robust}, PFs have been applied to a wide spectrum of signal-processing applications. Despite their advantages, the inherently high computational cost of PF limits their practical application, especially in real-time problems. This challenge has been addressed by algorithmic improvements~\cite{wang2009camshift}, efficient shared-memory~\cite{goodrum2012parallelization} and many-core~\cite{brown2012framework,hendeby2010particle,maskell2006single} implementations, and scalable distributed-memory solutions~\cite{zenker2010parallel,murray2013bayesian}. 

Here, we present a parallel software library for particle filtering, called the \texttt{PPF} library, which aims at providing application developers with an easy-to-use and scalable platform to develop PF-based parallel solutions for their applications. The main contributions of the \texttt{PPF} library are a highly optimized implementation and the extension of distributed resampling algorithms~\cite{Bolic2005} to hybrid shared-/distributed-memory systems. 

The library is written in Java and has an object-oriented architecture. It exploits a hybrid model of parallelism where the Message Passing Interface (MPI)~\cite{MPI} and Java threads are combined. The framework relies on the recent Java bindings of OpenMPI~\cite{OpenMPI,vega2013towards} for inter-process communication, and on Java threads for intra-process parallelism. Java is an emerging language in high-performance computing (HPC) offering new research opportunities. 

The \texttt{PPF} library includes implementations of different strategies for dynamic load balancing (DLB) across processes, and a checkerboard-like thread balancing scheme within processes. Inter-process DLB is used in the resampling phase of a PF, whereas thread balancing (i.e., intra-process balancing) is used throughout the entire library. Non-blocking point-to-point MPI operations are exploited wherever the chosen DLB strategy allows for them. Furthermore, the framework has interfaces for ImageJ~\cite{ImageJ}, Fiji~\cite{schindelin2012fiji}, and \texttt{imagescience}~\cite{Imagescience}, allowing these popular image-processing applications to directly access \texttt{PPF}'s application programming interface (API). The proposed framework can be used to facilitate implementation of other computationally demanding PF based techniques such as pMCMC~\cite{andrieu2010particle}, SMC$^{2}$~\cite{chopin2012smc2}, and SMS-(C)PHD~\cite{vo2003sequential}.

The manuscript is structured as follows: The generic PF algorithm is described in Section~2. In Section~3, distributed resampling algorithms are briefly discussed, followed by DLB schedules (Section~4) and effective particle routing (Section~5). In Section~6, we provide implementation details of the \texttt{PPF} library. Section~7 shows an application from biological imaging processing, where a single target is tracked through a sequence of 2D images. Two different distributed PF algorithms are compared. We summarize the contributions of the \texttt{PPF} library and discuss future work in Section~8. 

\section{Particle Filters}
\label{sec:sir} 
A generic PF algorithm consists of two parts: (i) sequential importance sampling (SIS) and (ii) resampling~\cite{Doucet2001}. A popular combined implementation of these two parts is the sequential importance resampling (SIR) algorithm~\cite{Doucet2001}. 
Recursive Bayesian importance sampling~\cite{Geweke1989} of an unobserved and discrete Markov process $\{\mathbf{x}_{k}\}_{k=1,\ldots ,K}$ is based on three components: (i) the measurement vector $\mathbf{Z}^k=\{\mathbf{z}_{1},\ldots ,\mathbf{z}_{k}\}$, (ii) the dynamics (i.e., state-transition) model, which is given by a probability distribution $p(\mathbf{x}_{k} | \mathbf{x}_{k-1})$, and (iii) the likelihood (i.e., observation model) $p(\mathbf{z}_{k} | \mathbf{x}_{k})$. Then, the state posterior $p(\mathbf{x}_{k} | \mathbf{Z}^{k})$ at time $k$ is recursively computed as:
\begin{equation}
\underset{\text{posterior}}{\underbrace{p(\mathbf{x}_{k} | \mathbf{Z}_{k})}} = \frac{\overset{\text{likelihood}} {\overbrace{ p(\mathbf{z}_{k} | \mathbf{x}_{k})}}\,\, \overset{\text{prior}} {\overbrace{p(\mathbf{x}_{k} | \mathbf{Z}^{k-1})}}}{\underset{\text{normalization}}{\underbrace{p(\mathbf{z}_{k} | \mathbf{Z}^{k-1})}}}\, ,
\end{equation}
where the prior is defined as:
\begin{equation}
p(\mathbf{x}_{k} | \mathbf{Z}^{k-1}) = \int p(\mathbf{x}_{k} | \mathbf{x}_{k-1}) \, p(\mathbf{x}_{k-1} | \mathbf{Z}^{k-1}) \, \mathrm{d}\mathbf{x}_{k-1}.
\end{equation}

Having the posterior, the
minimum mean square error (MMSE) or maximum {\it{a posteriori}} (MAP)
estimators of the state can be obtained~\cite{Doucet2001}, for example by \[\mathbf{x}_t^{\textrm{MMSE}}=\int \mathbf{x}_t p(\mathbf{x}_t|\mathbf{Z}^t)d\mathbf{x}_t.\]

Due to intractability of the high-dimensional integrals, PFs approximate the posterior at each time point $k$ by $N$ weighted samples (called ``particles'') $\{\mathbf{x}^i_k, w^i_k\}_{i=1,\ldots,N}$. This approximation amounts to Monte-Carlo quadrature and is achieved by sampling a set of particles from an importance function (proposal) $\pi(\cdot)$ and updating their weights according to the dynamics and observation models. This process is called sequential importance sampling (SIS)~\cite{Doucet2001}. SIS suffers from \textit{weight degeneracy} where small particle weights become successively smaller and do not contribute to the posterior any more. To overcome this problem, a \textit{resampling} step is performed~\cite{Doucet2001} whenever the number of particles with relatively high weights falls below a specified threshold. Through the standard notation~\cite{Bashi2003,Doucet2001}, the resulting SIR algorithm is given in Algorithm~\ref{alg:sir}.
To parallelize the SIR algorithm, one only needs to focus on the \textit{resampling} step, since all other parts of the SIR algorithm are local and can trivially be executed in parallel. 

\begin{algorithm}
\caption{SIR, a generic Particle Filtering Algorithm} \label{alg:sir}
\begin{algorithmic}[1]
\Procedure{SIR}{}
	\For{$i=1 \to N$} \Comment{Initialization, $k$=0} 
		\State $w_0^{i} \gets 1/N$
		\State Draw $\mathbf{x}_0^{i}$ from $\pi(\mathbf{x}_0)$
	\EndFor
	\For{$k=1 \to K$}  
		\For{$i=1 \to N$}  \Comment SIS step
			\State Draw a sample $\tilde{\mathbf{x}}_k^i$ from $\pi(\mathbf{x}_k | \mathbf{x}_{k-1}^i,\mathbf{Z}^{k})$
			\State Update the importance weights
			\State $\tilde{w}_k^i \gets w_{k-1}^i \frac{p(\mathbf{z}_{k} | \tilde{\mathbf{x}}_{k}^i)  p(\tilde{\mathbf{x}}_{k}^i | \mathbf{x}_{k-1}^i)}{\pi(\tilde{\mathbf{x}}_k^i | \mathbf{x}_{k-1}^i,\mathbf{Z}^{k})}$ 
		\EndFor
		\For{$i=1 \to N$} 
			\State $w_k^i \gets \tilde{w}_k^i / \sum_{j=1}^{N} \tilde{w}_k^j$
		\EndFor 
		
		\Comment Calculate the effective sample size
		\State $\widehat{N}_{\textrm{eff}} \gets 1 / \sum_{j=1}^{N} (w_{k}^{j})^2$ 
		\If{$\widehat{N}_{\textrm{eff}}<N_{\textrm{threshold}}$} \Comment Resampling step
				\State Sample a set of indices $\{s(i)\}_{i=1,\ldots ,N}$ distributed such that $\Pr[s(i)=l]=w_{k}^{l}$ for $l= 1 \to N$.
			\For{$i=1 \to N$}
				\State $\mathbf{x}_{k}^{i} \gets \tilde{\mathbf{x}}_{k}^{s(i)}$
				\State $w_{k}^{i} \gets 1/N$ \Comment Reset the weights
			\EndFor
		\EndIf
	\EndFor
\EndProcedure
\end{algorithmic}
\end{algorithm}

\section{Distributed resampling algorithms}
\label{sec:dra}
Distributed resampling algorithms (DRAs) can be categorized into three classes: The first includes \textit{multiple particle filter} (MPF) algorithms~\cite{strid2006parallel}, which are banks of truly independent PFs with no communication during and/or after local resampling. This renders MPF an embarrassingly parallel approach, in which every process runs a separate PF. The only communication in this case is sending the local estimates of the state to a master node that forms a combined estimate. However, this low communication complexity comes at the cost of a reduced statistical accuracy due to problems when, for example, some of the independent PFs diverge during the sequential estimation and introduce large errors into the combined estimate. If some of the independent PFs happen to explore the same area of state space, the approach is also computationally wasteful. Using a global resampling scheme, which incurs global communication between the processes, accuracy and efficiency can be improved.   

The second DRA category includes \textit{distributed resampling algorithms with non-proportional allocation} (RNA)~\cite{Bolic2005} and \textit{local selection} (LS) algorithms~\cite{Miguez2004}. These algorithms are designed to minimize inter-process communication. However, the way they perform resampling may cause a statistical accuracy loss, since the result generated by these algorithms depends on the number of processes used. A comparison of RNA with LS can be found in~\cite{Miguez2007}. In RNA, the state space is divided into disjoint strata, each of them assigned to a different process. The number of particles is fixed and the same for each process. The resampling step is performed locally by each process in its stratum, leading to an uneven weight distribution across processes. This requires deterministic particle routing (i.e., DLB) strategies. A popular choice is to migrate 10\%--50\% of the particles of each process to the neighboring process after resampling~\cite{Miguez2007,zenker2010parallel}. Neighborhood between processes is defined by a process topology, which usually is taken to be a ring. Using such a simple, static DLB scheme shortens application development times, but may cause redundant communication especially once the PF has converged onto its target and tracks it through state space. In this case, the PFs in all processes have converged to an equally good approximation of the posterior. The exchange of particles between  processes would then not be necessary anymore, but is still wastefully performed in RNA. It is also not clear what the optimal percentage of migrating particles should be. For applications requiring high precision, the number of particles that need to be communicated may also become too large, limiting the scalability of RNA. 

The third class of DRAs contains \textit{distributed resampling algorithms with proportional allocation} (RPA)~\cite{Bolic2005}. These algorithms are based on stratified sampling with proportional allocation, which means that particles with larger weights are resampled more often. This causes growing particle imbalance between the processes, since a process with higher-weight particles will generate more particles and thus become overloaded. Similarly, processes with low-weight particles will have even fewer particles after resampling and ``starve". To overcome this problem, RPA requires adaptive DLB schemes for particle routing, where the number of communication links is minimized in order to reduce the \textit{latency} in the network. 
In addition, also the sizes of the communicated messages have to be optimized in order to avoid \textit{bandwidth} problems. We address both the \textit{latency} and the \textit{bandwidth} criteria in RPA algorithms using the DLB schedules presented next. 

\section{DLB Schedules for RPA}
\label{sec:dlb}

To rebalance the workload (i.e., number of particles) on each process, one has to use a DLB scheme. \texttt{PPF} implements three DLB protocols, which all start by labeling the processes as either \textit{senders} or \textit{receivers}. A good DLB scheduler then minimizes the  communication overhead required for routing particles from the \textit{senders} to the \textit{receivers}. 

\subsection{Greedy Scheduler}
The Greedy Scheduler (GS) matches the first sender with the first receiver and then iterates through the senders. For each sender $S_i$ with particle surplus $N_{S_i}$, it moves as many particles as possible to receiver $R_j$. Once a receiver is full, it moves on to the next $R_{j+1}$ until the sender is empty. The procedure guarantees that at the end each process has the same number of particles. The pseudocode of GS is given in Algorithm~\ref{gs}.

\begin{algorithm}
\caption{Greedy Scheduler}\label{gs}
\begin{algorithmic}[1]
 \Require{S:=the list of senders, R:=the list of receivers.}
 \Ensure{schedule:=the list of matchings between the elements of S and R including the number of particles to be routed.}
\Procedure{GreedyScheduler}{$S,R$}
	\State {$j \gets 0$}
	\While{$S\neq \varnothing $}
		\While{$N_{S_i} \neq 0$}
			\If {$N_{S_i} \geq N_{R_j} > 0$}
				\State $\text{schedule} \gets \{S_i, R_j, N_{R_j}\}$ 
				\State $N_{S_i} \gets N_{S_i} - N_{R_j}$
				\State $N_{R_j} \gets 0$				
				\State $j \gets j + 1$
			\ElsIf {$N_{R_j}  > N_{S_i} \geq 0$} 
				\State $\text{schedule} \gets \{S_i, R_j, N_{S_i}\}$ 
				\State $N_{R_j} \gets N_{R_j} - N_{S_i}$
				\State $N_{S_i} \gets 0$								
				\State $j \gets 0$
			\EndIf
		\EndWhile
		\State $i \gets i + 1$
	\EndWhile
\State \textbf{return} schedule
\EndProcedure
\end{algorithmic}
\end{algorithm}

\subsection{Sorted Greedy Scheduler}
The Sorted Greedy Scheduler (SGS)~\cite{demirel2013balanced, demirel2013balancing,strid2006parallel} first sorts the senders in $S$ by their $N_{S_i}$ and the receivers in $R$ by their $N_{R_j}$, both in descending order. This sorting reduces the number of required communication links. The rest of the SGS algorithm is identical with GS, as seen in Algorithm~\ref{sgs}.

\begin{algorithm}
\caption{Sorted Greedy Scheduler}\label{sgs}
\begin{algorithmic}[1]
 \Require{S:=the list of senders, R:=the list of receivers.}
 \Ensure{schedule:=the list of matchings between the elements of S and R including the number of particles to be routed.}
\Procedure{SortedGreedyScheduler}{$S,R$}	
	\State $S' \gets$ sort($S$) \Comment{in descending order}
	\State $R' \gets$ sort($R$) \Comment{in descending order}
	\State \Return{ \Call{GreedyScheduler}{$S',R'$}}
\EndProcedure
\end{algorithmic}
\end{algorithm}

\subsection{Largest Gradient Scheduler}
While GS and SGS aim to balance the loads perfectly, one may be interested in a faster scheduler that causes less communication overhead, but does not guarantee optimal particle balancing. The Largest Gradient Scheduler (LGS) is such a sub-optimal heuristic. Similar to SGS, LGS first sorts $S$ and $R$ such that $N_{S_1} > N_{S_2} > \ldots > N_{S_{\left |  S \right |}}$ and $N_{R_1} > N_{R_2} > \ldots > N_{R_{\left | R \right |}}$. After that, each sender is paired with the corresponding receiver of same rank:
\begin{align*}
S_1 & \rightarrow R_1, \\
S_2  & \rightarrow R_2, \\
& \, \, \, \,  \vdots \\ 
S_{\min{(\left |  S \right |,\left |  R \right |)}} & \rightarrow R_{\min{(\left |  S \right |,\left |  R \right |)}}.
\end{align*}
LGS thus finds the \textit{largest gradients} between $S$ and $R$ and limits the number of communication links to
\begin{equation*}
 C=\min{(\left |  S \right |,\left |  R \right |)}.
 \end{equation*}
The pseudocode of LGS is given in Algorithm~\ref{lgs}.
\begin{algorithm}
\caption{Largest Gradient Scheduler}\label{lgs}
\begin{algorithmic}[1]
 \Require{S:=the list of senders, R:=the list of receivers.}
 \Ensure{schedule:=the list of matchings between the elements of S and R including the number of particles to be routed.}
\Procedure{LargestGradientScheduler}{$S,R$}	
	\State $S' \gets$ sort($S$) \Comment{in descending order}
	\State $R' \gets$ sort($R$) \Comment{in descending order}
	\For{$i = 1 \to \min{(\left |  S \right |,\left |  R \right |)}$}
		\If {$N_{S_i} \geq N_{R_i}$}
			\State $N_{S_i} \gets N_{S_i} - N_{R_i}$
			\State $\text{schedule} \gets \{S_i, R_i, N_{R_i}\}$  
		\Else
			\State $N_{S_i} \gets 0$
			\State $\text{schedule} \gets \{S_i, R_i, N_{S_i}\}$  
		\EndIf
	\EndFor
	\State \Return{schedule}
\EndProcedure
\end{algorithmic}
\end{algorithm}


%


\section{Parallel Particle Filtering Library}
\label{sec:ppff}
\begin{figure}[]
\centering
\includegraphics[width=0.6\columnwidth]{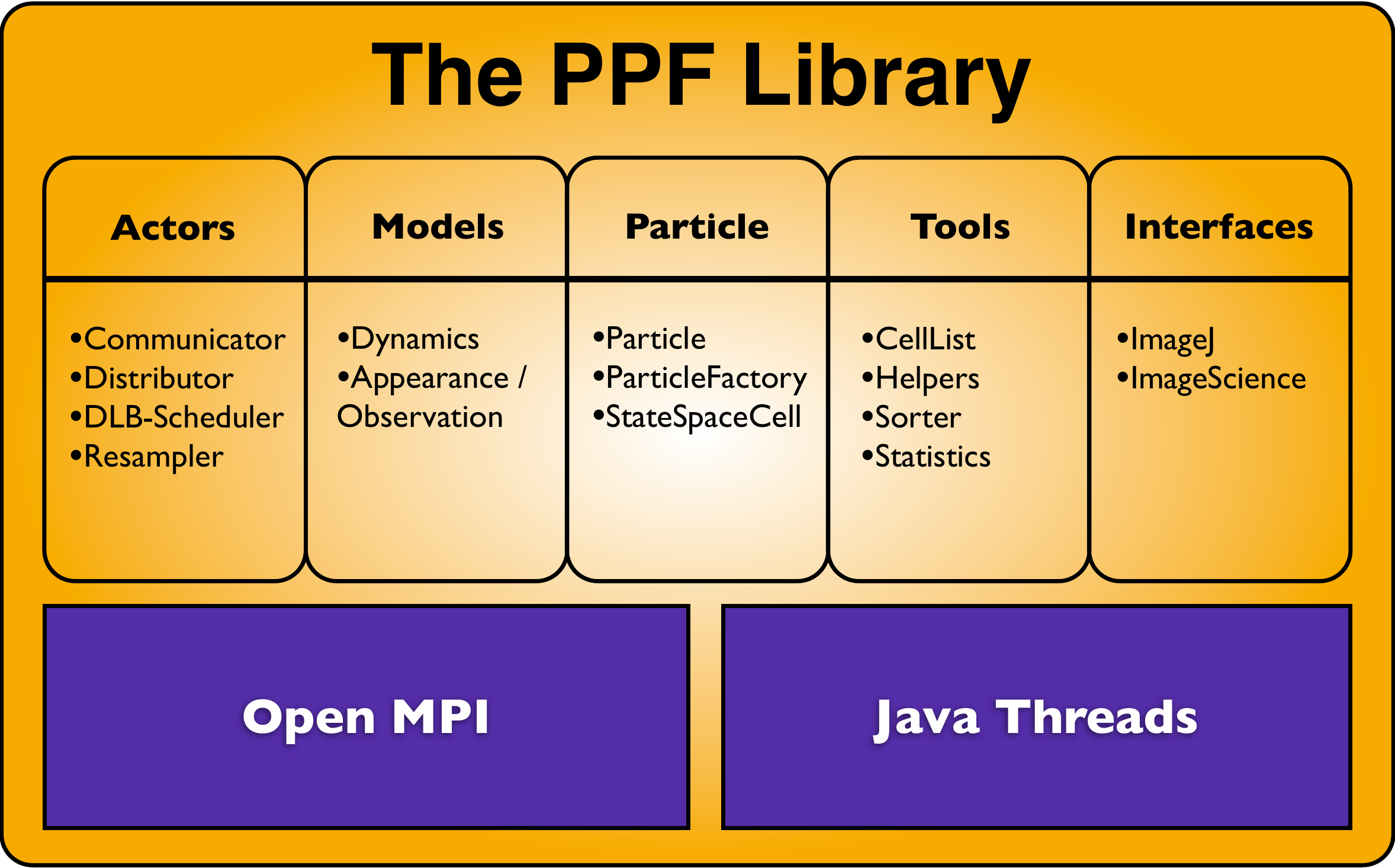}
\caption{Software structure of the \texttt{PPF} library. The library is divided into five modules and hides the complexity of MPI and multithreading from the application programmer. It also supports using \texttt{imagescience} classes, as well as functions from ImageJ and Fiji.}
\label{fig:ppfl}
\end{figure}

The \texttt{PPF} library is written in Java and uses the Message-Passing Interface (MPI)~\cite{MPI} for inter-process communication, which is the \textit{de facto} standard for parallel high-performance computing. The OpenMPI team~\cite{OpenMPI} recently started providing a Java interface based on \texttt{mpiJava}~\cite{Baker1999}, which covers all MPI 1.2 functions and is currently maintained on provisional basis in the development trunk of OpenMPI~\cite{OpenMPITrunk}. The \texttt{PPF} library also provides built-in support for ImageJ~\cite{ImageJ}, Fiji~\cite{schindelin2012fiji}, and \texttt{imagescience}~\cite{Imagescience} for file I/O, image analysis, and image editing. A schematic of the structure of the \texttt{PPF} library is shown in Fig.~\ref{fig:ppfl}. 

The library consists of five modules: (i) actors, (ii) models, (iii) particle, (iv) tools, and (v) interfaces. The \textit{actors} module encapsulates  functionality that is common to PF algorithms (e.g., resampling) and provides support for parallel PFs via its communication, data distribution, and DLB sub-modules. The \textit{models} module includes dynamics and observation models. By default, the library includes simple standard models that can be sub-classed to include application-specific models. 
The \textit{particle} module contains the particle data structure and related methods (e.g., particle generation). The \textit{tools} module contains a set of helper methods for sorting, statistical calculations, efficient particle neighbor lists, etc. The \textit{interfaces} module provides APIs to link the \texttt{PPF} library to ImageJ~\cite{ImageJ}, Fiji~\cite{schindelin2012fiji}, and \texttt{imagescience}~\cite{Imagescience}. This allows ImageJ/Fiji plugins to access the functionality provided by the \texttt{PPF} library, but it also allows \texttt{PPF} methods to use functions provided by ImageJ/Fiji, such as functions for image processing, file I/O, and graphical user-interface building. 

A client code that implements a parallel PF application can directly call the \texttt{PPF} API. Most of the intricacies arising from parallel programming and code optimization are hence hidden from the application programmer. Sitting as a middleware on top of MPI and Java threads, the \texttt{PPF} library makes the parallelization of PF applications on shared- and distributed-memory systems easier. Below, we highlight some of the features of the \texttt{PPF} library.

\subsection{Multi-level Hybrid Parallelism}
As HPC systems grow in size, multi-level hybrid parallelization techniques emerge as a viable tool to achieve high parallel efficiency in scalable scientific applications. Many applications~\cite{Lange2013} realize hybrid parallelization strategies by combining MPI with OpenMP~\cite{OpenMP}. In order to have full thread control and also enable job-level multi-threading, we here follow the trend of combining MPI for inter-process communication with native threads for intra-process parallelism. We employ Java's native thread concurrency model in the \texttt{PPF} library, which provides full thread control and avoids additional software/compiler installation and maintenance. Furthermore, we provide an intra-process load-balancing scheme for threads specifically designed for PF applications, whose implementation using Java threads is straightforward and easy to extend.

\begin{figure}[]
\centering
\includegraphics[width=0.5\columnwidth]{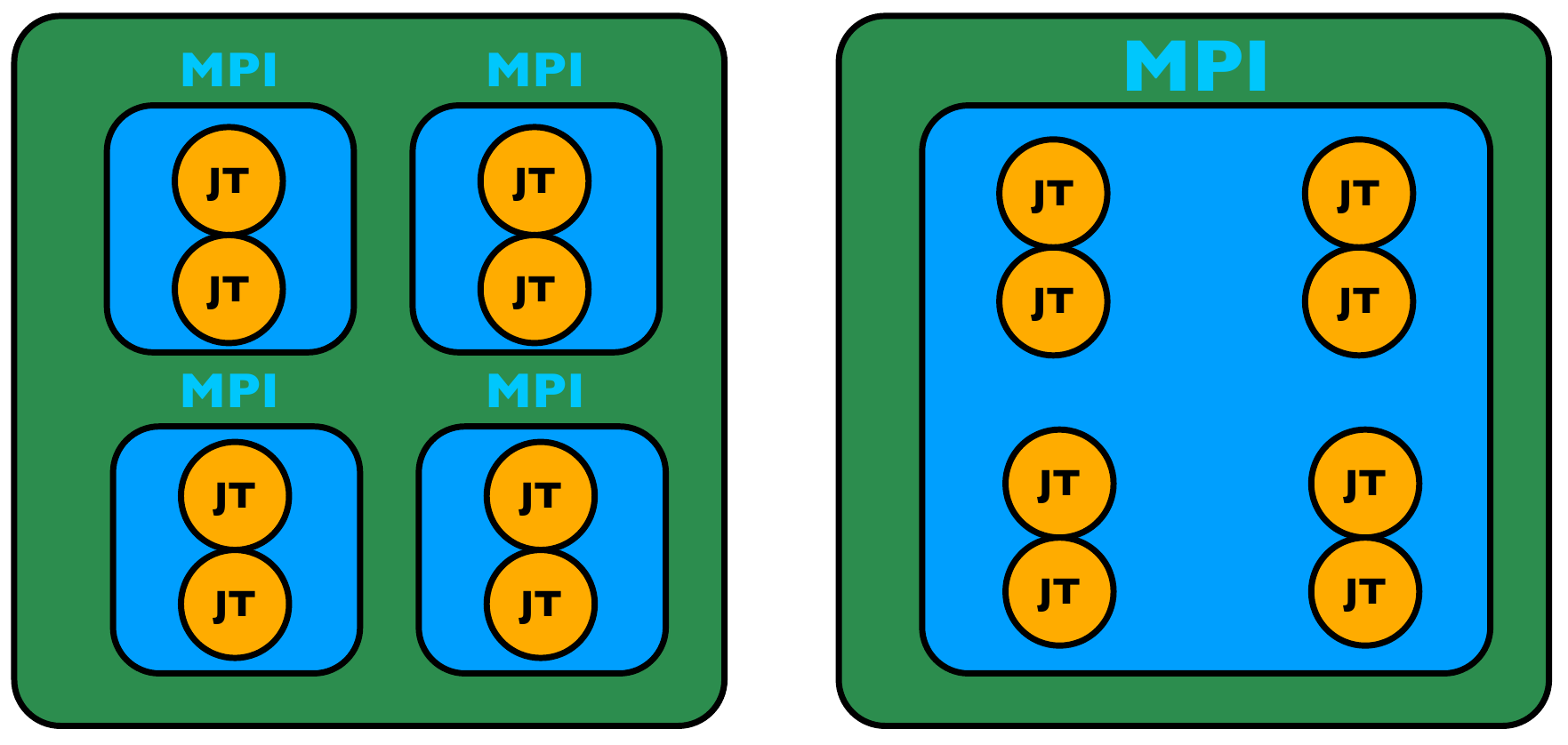}
\caption{Two possible ways to use hybrid parallelism with MPI and Java threads (JT) in \texttt{PPF}: On the left, each MPI process is bound to a \textit{CPU} (blue squares), and each JT is assigned to a \textit{core} (orange circles). On the right, there is one MPI process per \textit{node/computer} (green rectangle), and again one JT per \textit{core}. The \texttt{PPF} library implements both models and lets the application developer choose which one to use for a specific application.}
\label{hybrid}
\end{figure}

The \texttt{PPF} library lets the user choose between two different concurrency models, as illustrated in Fig.~\ref{hybrid}: In the first model (left panel), the number of MPI processes is equal to the number of available CPU chips, and the number of threads per process is equal to the number of cores per CPU. Compared to an all-MPI paradigm, this allows benefitting from shared caches between cores and reduces communication latency. 
The second model (right panel) uses a single MPI process per \textit{node/computer} and and one Java thread for each core. This hybrid model keeps the shared memory on the node coherent and causes a lower memory latency than an all-MPI implementation~\cite{Hybrid}. Which model one chooses for a particular application depends on the specific hardware (cache size, number of memory banks, memory bus speed, etc.) and application (data size, computational cost of likelihood evaluation, etc.). 

\subsection{Non-blocking MPI Operations}
During execution of DLB strategies, such as GS, SGS, or LGS, we use non-blocking point-to-point MPI operations in order to overlap communication with computation. This is especially useful during the DLB phase, where a sender sends its message to a receiver and then immediately carries on with generating new local particles. 

\subsection{Input-space Domain Decomposition}
In the \texttt{PPF} library, MPI-level parallelization is done at the particle-data level. This means that each MPI process has full knowledge of the input, e.g.~the image to be processed, but only knows a part of the particles in state space. At the thread level, we then also decompose the input (e.g., image) into smaller subdomains. This provides a convenient way of introducing thread-level parallelism. For images, the pixels containing particles are directly assigned to local threads. Thus, both the state space is distributed via MPI, and the input space is distributed across threads.

\subsection{Thread Balancing}
Thread-level load balancing is as important as process-level DLB. When using PFs for object tracking, for example, once an object is found and locked onto, many particles are drawn to the vicinity of that object. This worsens thread load-balance, since if consecutive pixels belong to the same thread, that thread becomes overloaded while others are idle. 

\begin{figure}[]
\centering
\includegraphics[width=0.5\columnwidth]{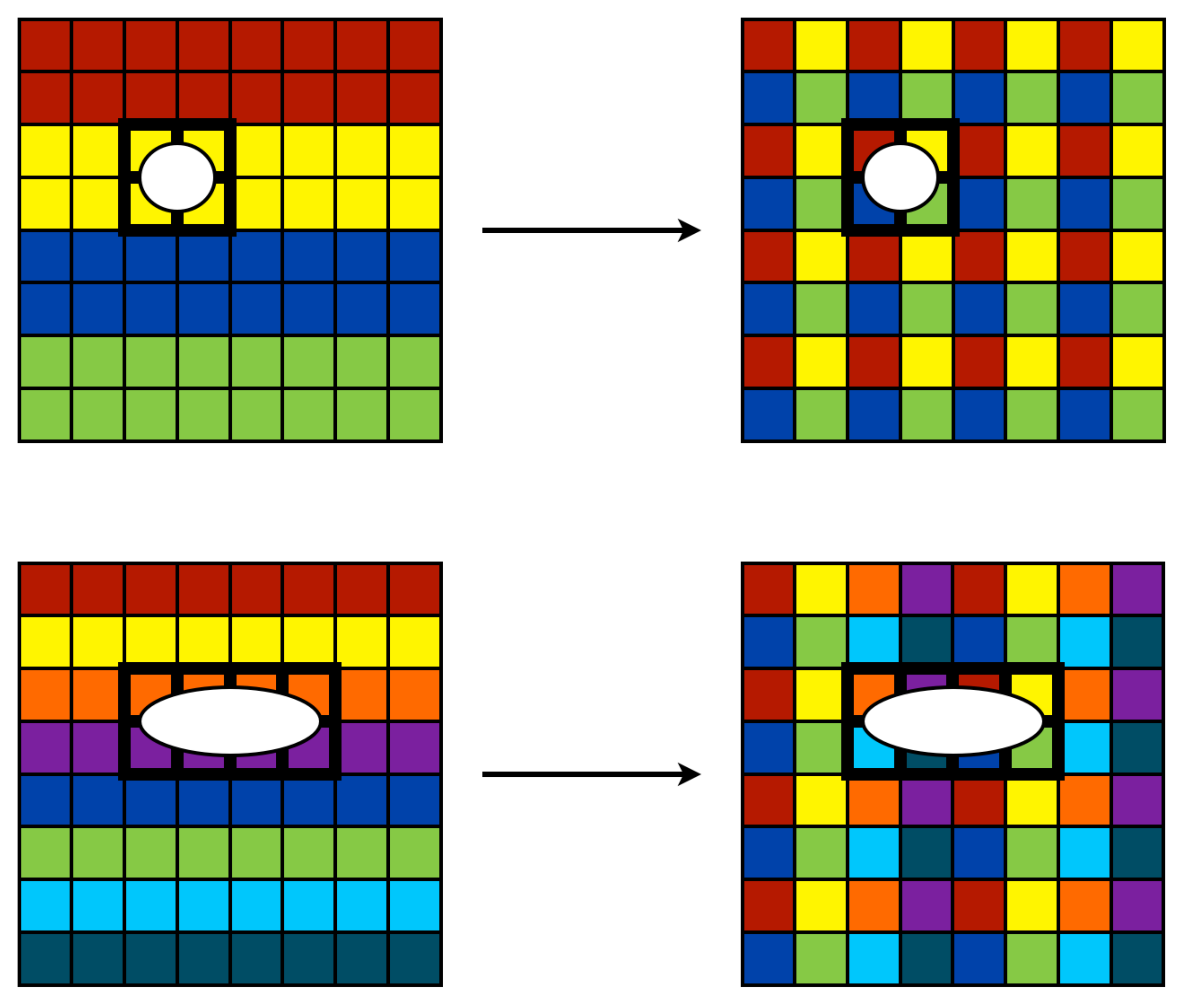}
\caption{Big boxes represent the binned state space (where in the case of images the pixels can be used as bins) and the colors represent which part of the state space will be processed by which thread. Once particles concentrate around an object to be tracked, i.e.~the posterior approximation converged, thread balancing accelerates the computations. The \texttt{PPF} library implements a checkerboard-like thread balancing scheme with patch sizes that depend on the support of the posterior distribution and on the number of threads. Two examples are shown here: In the \textit{2$\times$2} scheme (upper row), the area covered by the posterior (white circle) is distributed across four threads (upper right corner). Similarly, the \textit{2$\times$4} thread-balancing scheme (lower row) distributes a larger posterior (white ellipse) across eight threads. }
\label{thread_balancing}
\end{figure}

The \texttt{PPF} library hence uses an adaptive checkerboard-like load balancing scheme for threads, as illustrated in Fig.~\ref{thread_balancing}. Depending on the number of threads and the support of the posterior distribution, the size of the checkerboard patch is automatically adjusted. 

\subsection{Image Patches}
When using PFs for image processing, the likelihood computation involves image data and may be computationally costly due to, e.g., invoking a  numerical simulation of the image-formation process. Performance improvements in the likelihood calculation hence have the largest impact on the overall performance of a PF image-processing application. 

In many image-based applications, a separate likelihood estimation is carried out for each particle, where the full image is loaded and then the likelihood kernel is applied. In image-based likelihood computations, these kernels are typically symmetric (e.g., Gaussian) and local (i.e., span only a few pixels). Thus, it is sufficient to visit pixels one by one and only load the \textit{image patch} centered at the visited pixel. The size of the patch is given by the likelihood kernel support, which is typically much smaller than the whole image. Once a patch is loaded, computing the particle weights in the central pixel requires less time. In fact, if there are $N_\text{pix}$ pixels in the image and $N$ particles in state space, the overall computational complexity of the likelihood calculation is reduced from $\mathcal{O}(N N_\text{pix})$ to 
$\mathcal{O}(N)$, which is usually a significant reduction. 

Since threads are distributed in a checkerboard-like fashion, only one thread needs to load an image patch and all neighboring threads can simply access the patch data from shared cache. This results in better cache efficiency. 

\subsection{Piecewise Constant Sequential Importance Resampling}
The \texttt{PPF} library also provides an implementation of a fast approximate SIR algorithm that uses a piecewise constant approximation of the likelihood function to estimate the posterior distribution faster. This algorithm is called Piecewise Constant Sequential Importance Sampling (pcSIR) and can offer significant speedups~\cite{demirel2014piecewise}. 

\section{An Example Application}
\label{sec:application}
We demonstrate the capabilities of the \texttt{PPF} library by using it to implement a PF application for tracking sub-cellular objects imaged by fluorescence microscopy~\cite{akhmanova2005,komarova2009mammalian}. In this example from biological microscopy imaging, sub-cellular structures such as endosomes, vesicles, mitochondria, or viruses are fluorescently labeled and imaged over time with a microscope. Many biological studies start from analyzing the dynamics of those structures and extracting parameters that characterize their behavior, such as average velocity, instantaneous velocity, spatial distribution, motion correlations, etc. First, we describe the dynamics and appearance models implemented in the \texttt{PPF} library for this biological application, and then we explain the technical details of the experimental setup. 

\subsection{Dynamics Model}
The motion of sub-cellular objects can be represented using a variety of motion models, ranging from random walks to constant-velocity models to more complex dynamics where switching between motion types occurs~\cite{godinez2012identifying,smal_media}. 

Here, we use a near-constant-velocity model, which is frequently used in practice~\cite{smaltmi,rong2003survey}. The state vector in this case is $\mathbf{x}=(\hat{x}, \hat{y}, v_x, v_y, I_0)^T$, where $\hat{x}$ and $\hat{y}$ are the estimated $x$- and $y$-positions of the object, $(v_x,v_y)$ its velocity vector, and $I_0$ its estimated fluorescence intensity. 

\subsection{Observation Model}
Many sub-cellular objects are smaller than what can be resolved by a visible-light microscope, making them appear in a fluorescence image as diffraction-limited bright spots, where the intensity profile is given by the impulse-response function of the microscope, the so-called point-spread function (PSF)~\cite{smaltmi, smal_media,helmuth2010beyond,helmuth2009deconvolving}. 

In practice, the PSF of a fluorescence microscope is well approximated by a 2D Gaussian~\cite{thomann2002automatic}. 
The object appearance in a 2D image is hence modeled as:
\begin{equation}
I(x,y;x_0, y_0) = I_0 \exp\left(-\frac{(x-x_0)^2 + (y-y_0)^2}{2\sigma^2_{\textrm{PSF}}}\right) + I_{\textrm{bg}},
\end{equation}     
where $(x_0, y_0)$ is the position of the object, $I_0$ is its intensity, $I_{\textrm{bg}}$ is the background intensity, and $\sigma_{\textrm{PSF}}$ is the standard deviation of the Gaussian PSF. Typical microscope cameras yield images with pixel edge lengths corresponding to 60 to 200\,nm physical length in the specimen. For the images used here, the pixel size is 67\,nm, and the microscope has $\sigma_{\textrm{PSF}}=78$\,nm (or 1.16 pixels). During image acquisition, the ``ideal'' intensity profile $I(x,y)$ is corrupted by measurement noise, which in the case of fluorescence microscopy has mixed Gaussian-Poisson statistics. For the resulting noisy image $\mathbf{z}_k=Z_k(x,y)$ at time point $k$, the likelihood $p(\mathbf{z}_k|\mathbf{x}_{k})$ is:
\begin{equation} 
p(\mathbf{z}_k|\mathbf{x}_{k}) \varpropto \exp\!\!\left(\!-\frac{1}{2\sigma^2_{\xi}}\!\!\sum_{(x_i, y_i)\in\mathcal{S}_{\mathbf{x}}}\!\!\!\!\!\!\left [Z_k(x_i, y_i)-I(x_i, y_i;\hat{x}, \hat{y})\right ]^2\!\!\right)\!\!,
\end{equation}     
where $\sigma_{\xi}$ controls the peakiness of the likelihood, $(x_i, y_i)$ are the integer coordinates of the pixels in the image, $(\hat{x}, \hat{y})$ are the spatial components of the state vector $\mathbf{x}_k$, and $\mathcal{S}_{\mathbf{x}}$ defines a small region in the image centered at the location specified by the state vector $\mathbf{x}_k$. Here, $\mathcal{S}_{\mathbf{x}}=[\hat{x} - 3\sigma_{\textrm{PSF}}, \hat{x} + 3\sigma_{\textrm{PSF}}]\times[\hat{y} - 3\sigma_{\textrm{PSF}}, \hat{y} + 3\sigma_{\textrm{PSF}}]$. 

\begin{figure}[]
\centering
\includegraphics[width=0.8\columnwidth]{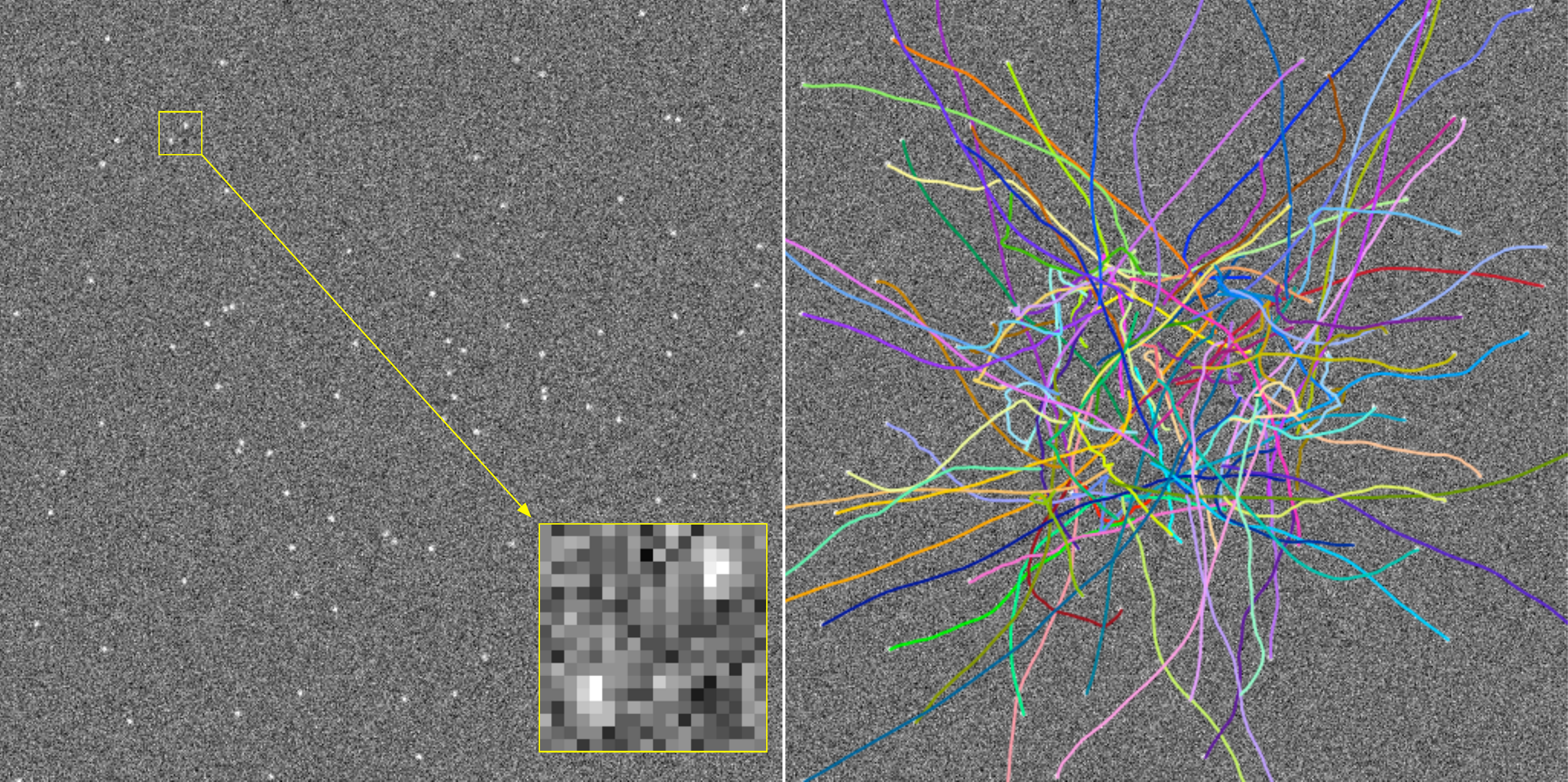}
\caption{Examples of low SNR synthetic images used in the experiments. Left: The first $512\times 512$ frame from a movie sequence of 50 frames, showing the typical object appearance due to the Gaussian PSF model. Right: Trajectories of the moving bright objects, generated using the nearly-constant-velocity dynamics model, overlaid with the first image frame.}
\label{syndata}
\end{figure}

\subsection{Experimental Setup}
We consider a single-object tracking problem where $\sigma_{\textrm{PSF}}=1.16$ pixels and the images have a signal-to-noise ratio (SNR) of 2 (equivalent to 6 dB). An example of an input image is shown in Fig.~\ref{syndata} (left). It is a synthetic image showing a number of bright PSF spots moving according to the above-described dynamics and appearance models. The task considered here for the PF is to detect the spots and track their motion over time, reconstructing their trajectories. The ground-truth trajectories used to generate the synthetic movies are shown in Fig.~\ref{syndata} (right). Comparing PF tracking results with them allows quantifying the tracking error.
For other applications, the \texttt{PPF} library can easily be extended to include also other dynamics and observation models.

Using double-precision arithmetics, a single particle requires 52\,B (i.e., six doubles and one integer) of computer memory. The particles are initialized uniformly at random and all tests are repeated 20 times with different random synthetic movies on the MadMax computer cluster of MPI-CBG, which consists of 44 nodes each having two 6-core Intel\textregistered \,  Xeon\textregistered \, E5-2640 2.5\,GHz CPU with 128\,GB DDR3 800\,MHz memory. All algorithms are implemented in Java (v.~1.7.0\_13), and OpenMPI (v.~1.9a1r28750) is used for inter-process communication.

We test the \texttt{PPF} library using both the RNA and RPA algorithms with different problem sizes and computer system sizes up to 384 cores.

\subsection{Results with RNA}
In RNA, each processor maintains the same amount of particles and thus a DLB scheme is not required. The \texttt{PPF} library implements both the classical RNA as well as \textit{adaptive RNA} (ARNA)~\cite{demirel2014adaptive}. ARNA dynamically adjusts the particle-exchange ratio by keeping track of the number of processes actually involved in successfully tracking an object. These processes are then arranged in a ring topology, such that  particle routing is straightforward. In order to find the tracked object again if it has been lost for a couple of frames, ARNA randomizes the process order in the ring topology whenever the object is lost, thus ensuring faster information exchange to re-locate the object. 

\begin{figure}[]
\centering
\includegraphics[width=0.6\columnwidth]{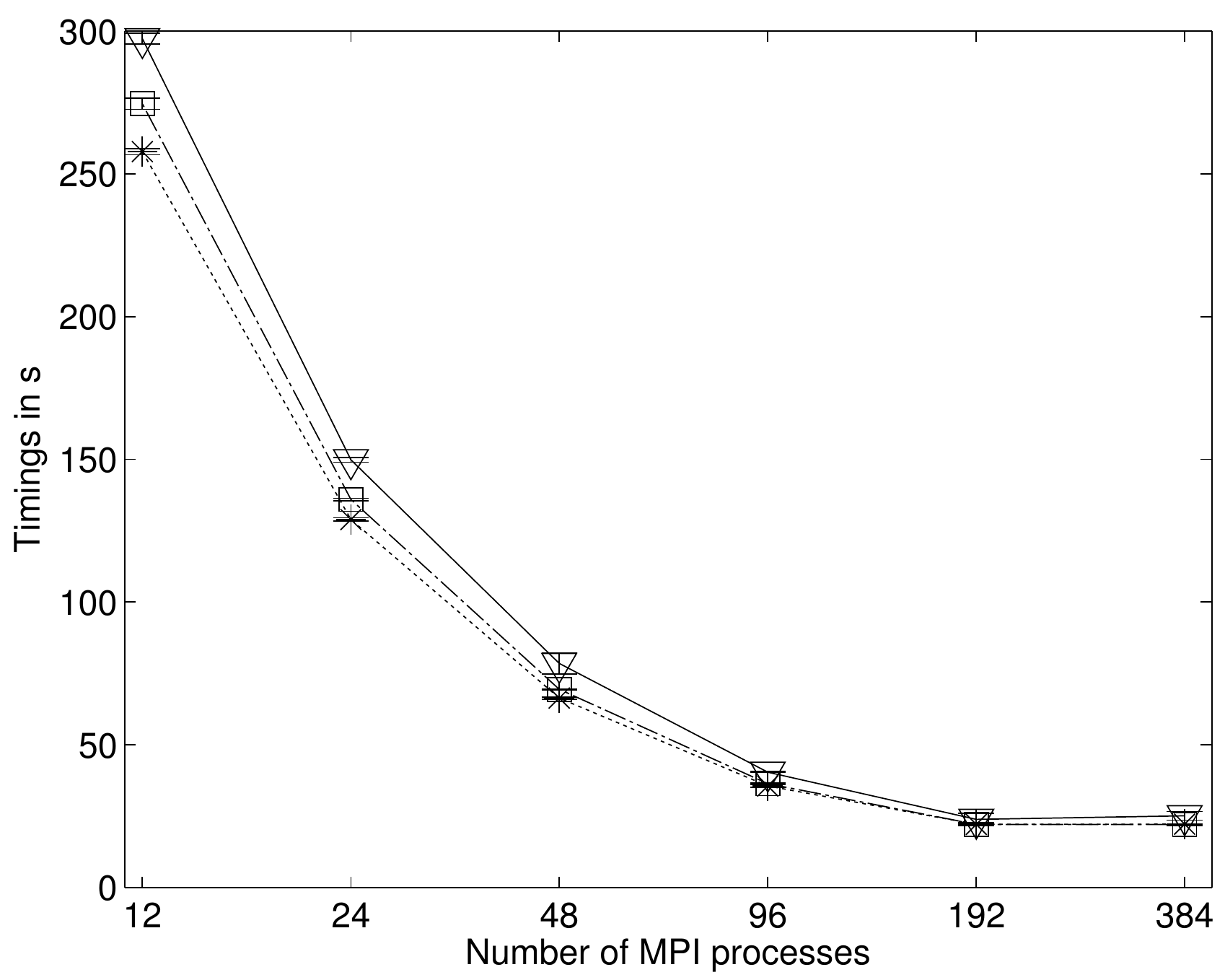}
\caption{Absolute wall-clock times of RNA with 10\% ($\square$) and 50\% ($\triangledown$) particle-exchange ratios and of ARNA ($\star$) for a strong scaling with 38.4 million particles. This problem size allows efficient parallelism on up to 192 cores.}
\label{fig:strong1threadrna}
\end{figure}

\begin{figure}[]
\centering
\includegraphics[width=0.6\columnwidth]{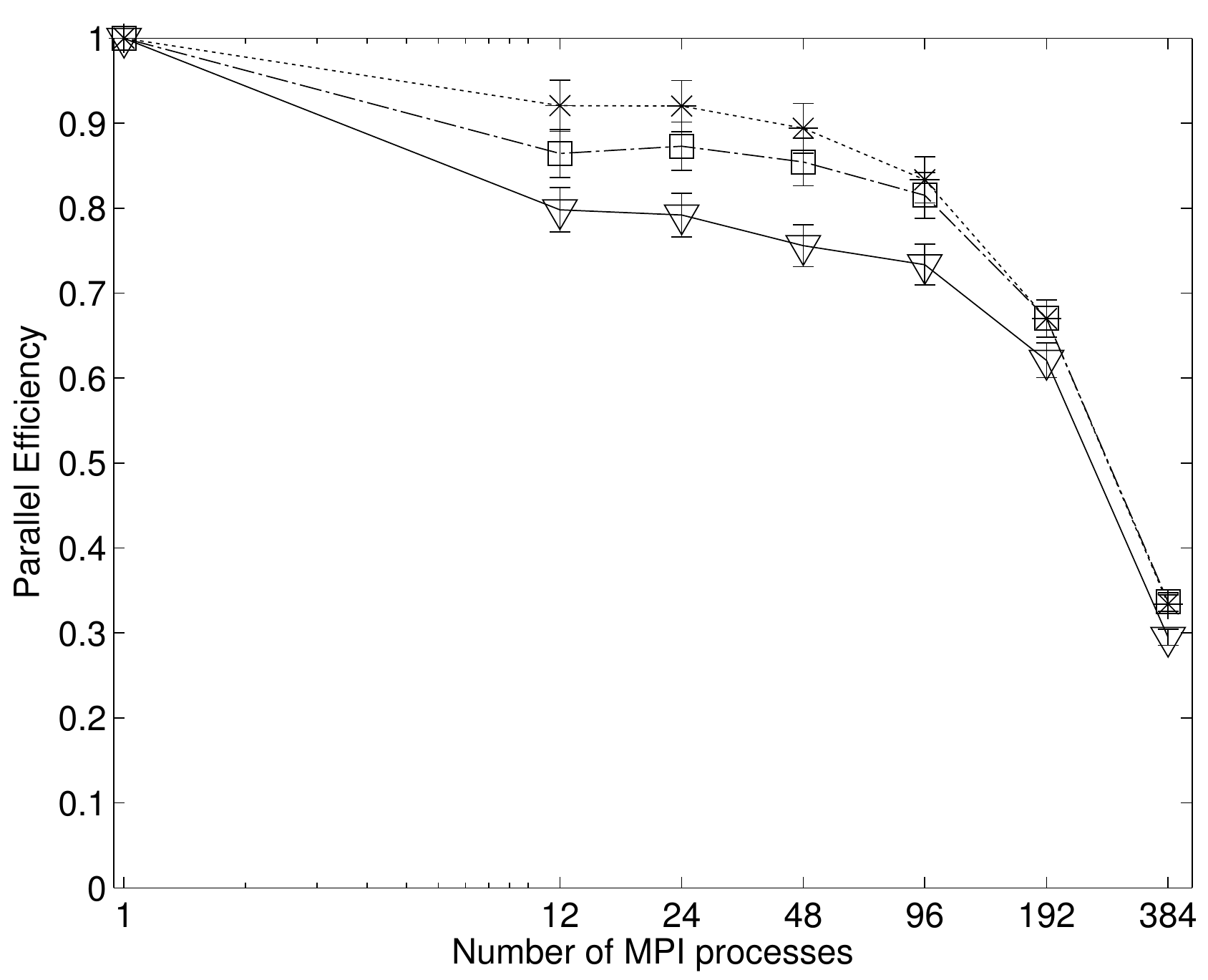}
\caption{Parallel efficiencies of RNA with 10\% ($\square$) and 50\% ($\triangledown$) particle-exchange ratios and of ARNA ($\star$) for a strong scaling with 38.4 million particles. This problem size allows efficient parallelism on up to 192 cores.}
\label{fig:rnaparefficiency}
\end{figure}

In the example described above, the \texttt{PPF} library reduces the runtime from about 50 minutes on a single core to about 20 seconds on 192 cores.
Figure~\ref{fig:strong1threadrna} shows the wall-clock runtimes for a strong scaling with a constant number of 38.4 million particles (1.86 GB of particle data) distributed across an increasing number of cores. RNA and ARNA show parallel efficiencies of 65\% and 67\%, respectively, on 192 cores for the selected problem size (Fig.~\ref{fig:rnaparefficiency}). Beyond 192 cores, the parallel efficiencies drop below 40\% for all RNA variants, as the number of particles per process becomes too small to amortize the communication overhead. For all RNA simulations, we use a hybrid parallelism model defined in Fig.~\ref{hybrid} (left) where each MPI process is assigned a single Java Thread. 

\subsection{Results with RPA}
\begin{figure}[]
\centering
\includegraphics[width=0.55\columnwidth]{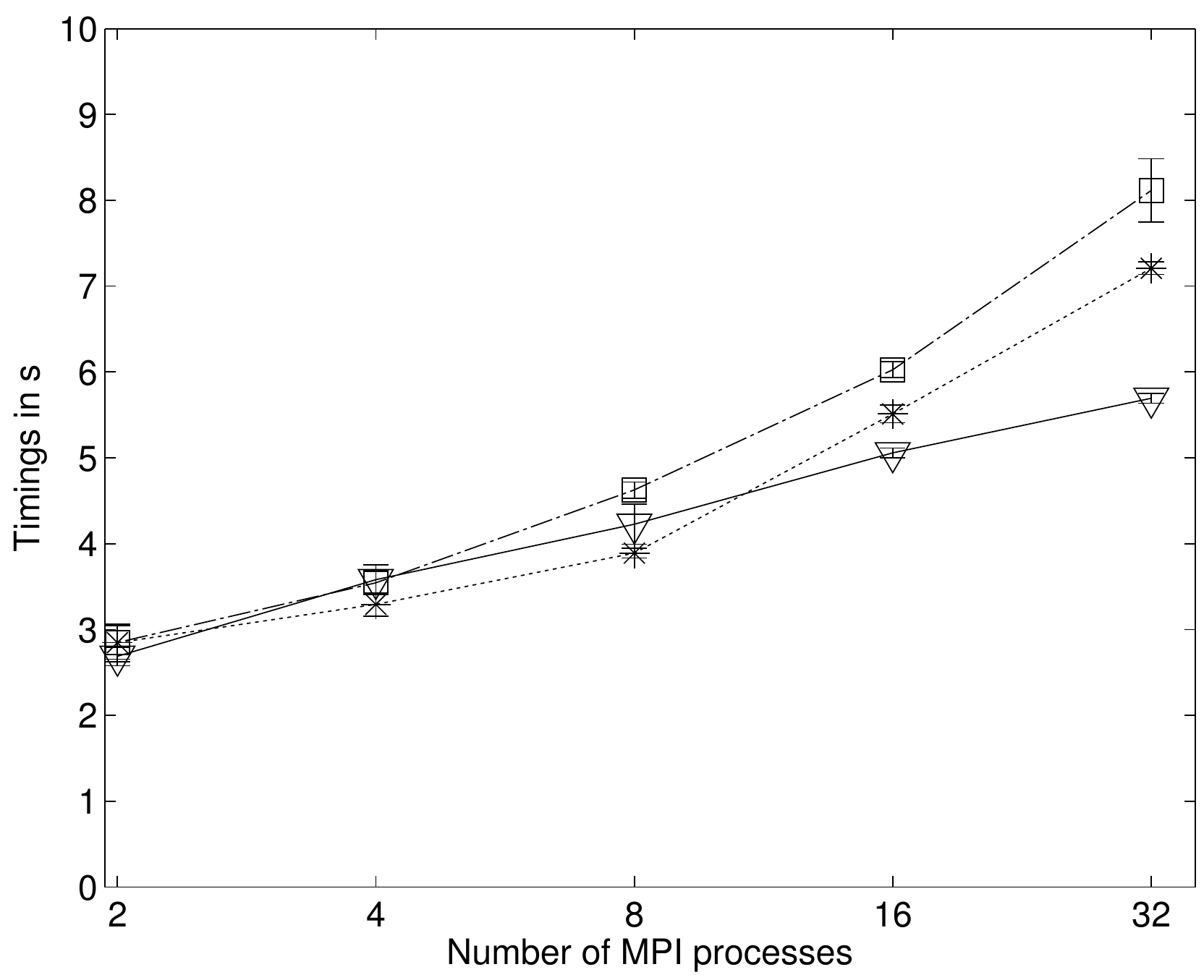}
\caption{Weak scaling runtime results with less than a second standard deviation of a RPA run with 60k particles per MPI process. Each MPI processes is mapped onto a single CPU with six logical cores. There are six Java threads per MPI process. Three DLB schemes are used: Greedy ($\square$), SortedGreedy ($\star$), and LGS ($\triangledown$).}
\label{fig:weakrpa6thread}
\end{figure}

\begin{figure}[]
\centering
\includegraphics[width=0.5\columnwidth]{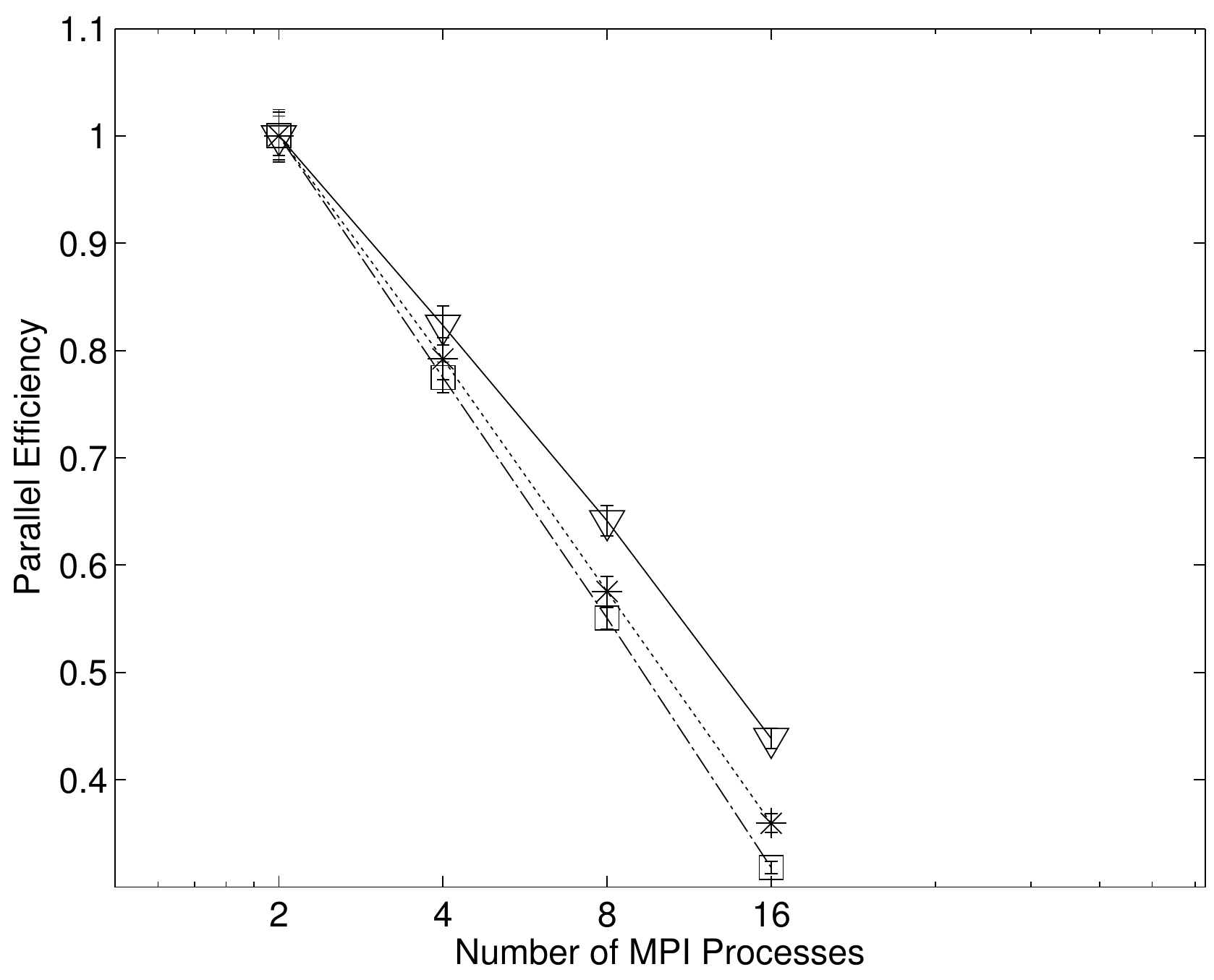}
\caption{Strong-scaling parallel efficiencies of RPA with a constant number of 3.84 million particles distributed across an increasing number of processes. 
Three different DLB schemes are compared: Greedy ($\square$), SortedGreedy ($\star$), and LGS ($\triangledown$). Each MPI process is pinned to one of the two available CPUs on each node, each running six Java threads.}
\label{fig:parEffRPA}
\end{figure}

For RPA, we compare three different DLB schemes. The tracking accuracy is measured by the root-mean-square error (RMSE) and was the same for all tests (about 0.063 pixels). All DLB schemes hence lead to results of equal quality. We use six Java threads per MPI process, since each CPU of the benchmark machine has six cores, and one MPI process per CPU. The wall-clock times are shown in Fig.~\ref{fig:weakrpa6thread} for a weak scaling with 60\,000 particles per process. The corresponding parallel efficiency is shown in Fig.~\ref{fig:parEffRPA}. Overall, LGS provides the best scalability, due to its linear communication complexity. Nevertheless, RPA scales less well than RNA and ARNA. For all RPA tests, we use a hybrid parallelism model defined in Fig.~\ref{hybrid} (right) where each MPI process is assigned six Java Threads. 

\section{Conclusions}\label{sec:concl}
We presented the \texttt{PPF} library that enables parallel particle filtering applications on commodity as well as on high-performance parallel computing systems. The library uses multi-level hybrid parallelism combining OpenMPI with native Java threads. As a proof of concept for the recently developed OpenMPI Java bindings, we showed the capability of the \texttt{PPF} library in a biological imaging application. The \texttt{PPF} library reduces parallel runtimes of the RNA and RPA methods by integrating dynamic load balancing with thread balancing and also implements PF-specific algorithmic improvements such as domain decomposition, image patches, and piecewise constant sequential importance resampling (pcSIR). The \texttt{PPF} library renders using parallel computer systems easier for application developers by hiding the intricacies of parallel programming and providing a simple API to design parallel PF applications. We presented a test case where 1.86\,GB of particle data were distributed across 192 cores at 67\% efficiency with either the RNA or ARNA methods used.

Future work includes adding support for graphics processing units (GPU) to further accelerate certain parts of PF algorithms and designing new DLB algorithms for RPA that exploit MPI 3.0 extensions, such as non-blocking collective operations. 

The \texttt{PPF} library is available as open source and free of charge from the MOSAIC Group's web page \url{http://mosaic.mpi-cbg.de}, under the ``Downloads'' section.

\section*{Acknowledgments}
The authors thank the MOSAIC Group (MPI-CBG, Dresden) for fruitful discussions and the MadMax cluster team (MPI-CBG, Dresden) for operational support. \"{O}mer Demirel was funded by grant \#200021--132064 from the Swiss National Science Foundation (SNSF), awarded to I.F.S. Ihor Smal was funded by a VENI grant (\#639.021.128) from the Netherlands Organization for Scientific Research (NWO).

\bibliographystyle{unsrt}
\bibliography{particle_filtering}

\end{document}